# Experimentally observed defect tolerance in the electronic structure of lead bromide perovskites

Gabriel J. Man[1,2,3*], Aleksandr Kalinko[4], Dibya Phuyal[5], Pabitra K. Nayak[6], Håkan Rensmo[1], Sergei M. Butorin[1*]

**Author addresses:**

1. Condensed Matter Physics of Energy Materials, Division of X-ray Photon Science, Department of Physics and Astronomy, Uppsala University, Box 516, Uppsala 75121, Sweden
2. GJM Scientific Consulting, Fort Lee, New Jersey 07024, USA
3. MAX IV Laboratory, Lund University, PO Box 118, Lund 22100, Sweden
4. Deutsches Elektronen-Synchrotron DESY, Notkestraße 85, Hamburg 22607, Germany
5. Division of Material and Nano Physics, Department of Applied Physics, KTH Royal Institute of Technology, Stockholm 10691, Sweden
6. Tata Institute of Fundamental Research, 36/P, Gopanpally Village, Serilingampally Mandal, Hyderabad 500046, India

* Correspondence and requests for materials addressed to gabriel.man@maxiv.lu.se and sergei.butorin@physics.uu.se.

**Abstract**

Point defect tolerance in materials, which extends operational lifetime, is essential for societal sustainability, and the creation of a framework to design such properties is a grand challenge in the material sciences. Using three prototypical lead bromide perovskites in single crystal form and high-resolution synchrotron-based X-ray spectroscopy, we reveal the unexpectedly pivotal role of the A-cation in mediating the influence of photoinduced defects. Organic A-cation hydrogen bonding facilitates chemical flexing of the lead-bromide bond that mitigates the self-doping effect of bromide vacancies. The contribution of partially ionic lead-bromide bonding to the electronic band edges, where the bonding becomes more ionic upon the formation of defects, mitigates re-hybridization of the electronic structure upon degradation. These findings reveal two new general design principles for defect tolerance in materials. Our findings uncover the foundations of defect tolerance in halide perovskites and have implications for defect calculations, all beam-based measurements of photophysical properties and perovskite solar cell technology.

## INTRODUCTION

The vision of sustainable human society has renewed research interest in defect tolerant and/or autonomously self-healing materials that may yield longer operational lifetimes for devices. Such materials include defect tolerant solar absorbing semiconductors which were first investigated over two decades ago [1–4]. Indeed, the use of such materials has been argued as essential for sustainability since 100% recycling of materials is near impossible due to entropy [5]. Defect tolerance in general consists of a potentially incomplete set of five material design principles [1]. For photovoltaic materials it is defined as a property where "the extrinsic, intrinsic, or structural defects that do form have a very minimal effect on mobility (μ) and minority-carrier lifetime (τ)" [3]. Lead halide perovskites (HaP) of the form $APbX_3$ have attracted substantial renewed research interest for over a decade, motivated by initially dramatic gains in HaP solar cell efficiencies and now other optoelectronic applications [6–10]. The attractive device-level properties of HaPs are partly attributed to the purported tolerance of the HaP electronic structure against structural and compositional defects [11]. This is rationalized by the different characters of the states at the valence band maximum (VBM) and conduction band minimum (CBM); for example, halide *p*-lead *6s* anti-bonding states at the VBM and halide *p*-lead *6p* anti-bonding states at the CBM [3,12,13]. The energy levels of compositional/structural defects are expected to be located within the VB or near the CBM (shallow defect) as opposed to deeper within the bandgap, which could facilitate nonradiative recombination of photogenerated carriers. Defect tolerance in the electronic structure of HaPs is a generally accepted concept though the origin(s) are controversial [14].

Challenges remain in verifying the existence of structural/compositional defect tolerance of the electronic structure of HaPs and, if present, precisely defining its origins. If defect tolerance indeed exists in HaPs, then defects may be electrically and optically inactive, which poses challenges for their detection. We note that defect tolerance can refer to the (in)effectiveness of nonradiative carrier capture by defects (affects minority-carrier lifetime and mobility) or the (non)modification of the electronic structure in the presence of defects (affects carrier mobility), and here we focus on electronic structure. Intrinsic crystal structure complexity (polymorphism) in HaPs has been reported, and work is ongoing to manage polymorph transitions which impact optoelectronic properties such as the bandgap and photoactivity [15–18]. Defect-related polymorphism further complicates the study of defect tolerance as it adds the requirement to differentiate between electronic structure changes due to intrinsic versus defect-related polymorphism. Computation with density functional theory (DFT) has mostly led the way thus far but the quantities calculated, such as the crystal structures of cation- and/or anion-deficient phases, defect energy levels, etc. have been found to depend sensitively on the choice of DFT functional, incorporation of structural dynamics, etc. [11,19–21]. Hence, experimental support is urgently required but progress has been hindered by variations in sample quality, material susceptibility to beam damage and more [22–24].

Here we use element-selective and bulk-sensitive, High Energy Resolution Fluorescence Detected X-ray Absorption Spectroscopy (HERFD-XAS) and X-ray emission spectroscopy (XES) measurements of three prototypical lead bromide perovskites, in single crystal form, to concurrently assess potential changes to the lead-bromide (Pb-Br) bonding, (A-cation)-bromide (A-Br) bonding, $PbBr_6$ octahedral structure and bromine- and lead-projected density-of-states (Br PDOS, Pb PDOS) in the electronic structure during controlled photodegradation. Our work reveals the pivotal role of the A-cation in mediating bromide vacancy interaction with the lead-bromide sublattice in a regime that does not involve degradation into $PbBr_2$ and ABr. The knowledge gained from our findings has confirmed and extended the set of general design principles for defect tolerant materials.

## RESULTS

We utilize controlled photodegradation conditions and concurrent measurements of the Pb-Br bond ionicity, A-Br bonding, $PbBr_6$ structure and bromine- and lead-PDOS to uncover A-cation-influenced material changes due to photodegradation. Further details on the types of information yielded by bromine *K*-edge and lead $L_3$-edge XAS and bromine valence-to-core (VtC) XES are found in Supplementary Note 1. The lead bromide perovskite (APB) subfamily is an optimal platform for isolating A-cation influence on defect tolerance as the three prototypical APB (A = methylammonium (MA), formamidinium (FA), cesium (Cs)) are all thermodynamically stable at room temperature, and is technologically relevant [25,26]. While the lead iodide perovskite (API) subfamily possesses bandgaps better suited for single-junction solar energy conversion, cesium lead tri-iodide (CsPI) and the photoactive phase of formamidinium lead tri-iodide (FAPI) are thermodynamically unstable at room temperature [27,28]. We exploit the susceptibility of halide perovskites to photon beam damage by simultaneously measuring on and photodegrading the materials [22,23]. The use of high-quality single crystals enables us to start with an initial, well-defined set of polytypes; our recent work with similar crystals and measurements showed close agreement between the measured and calculated bromine *K*-edge XAS [29]. Photodegradation with X-rays and nonionizing visible light are expected to yield similar degradation mechanisms. The halide ion has been found to be the most mobile ionic defect and been implicated in the degradation of perovskite solar cells [30,31]. We find direct (σ states in bromine *K* spectra) and indirect (structural information from lead $L_3$ spectra) evidence for bromide vacancy formation, which is expected to be accompanied by bromide ion formation. With a measured X-ray spot size of ~220 µm x 100 µm, probing depth in excess of 10 µm and single crystal cuboid length of ~2 mm, surface effects arising from photodegradation can be reasonably excluded from our measurements. The measurements were performed in a dry nitrogen atmosphere to exclude the effects of air and humidity.

The measured bromine *K* HERFD spectra, lead $L_3$ HERFD spectra and bromine VtC spectra of pristine and photodegraded FAPB, MAPB and CsPB are presented in Figure 1. The spectra of the pristine APB compounds are similar to previously reported spectra [29]. The pristine and photodegraded spectra were recorded from the same, initially fresh spot, and the photodegraded spectra were recorded at least ~110 minutes after the recording of the pristine spectra from the same spot, with continuous beam exposure. The bromine $K\beta_1$ XES of $PbBr_2$ and pristine and photodegraded FAPB, MAPB and CsPB are displayed in Supplementary Figure 1; the $K\beta_1$ XES of the pristine APB are similar to reported spectra [29]. During the experiment, we visibly observed that the X-ray excited optical luminescence (XEOL) was green (consistent with the ~2.3 eV bandgap of the APB's) and both the color and intensity remained roughly constant during the entire course of photodegradation. To ascertain whether the dominant photodegradation mechanism involves decomposition of the organic $APbBr_3 \rightarrow ABr + PbBr_2$, we have performed four spectroscopy-based checks (Supplementary Note 2, involving Supplementary Fig. 1-4) and conclude that $APbBr_3 \rightarrow ABr + PbBr_2$ decomposition, if present, is negligible in the sample regions probed and the photodegraded APB's remain perovskites. This conclusion is consistent with our visual observation of the XEOL color and intensity being roughly constant during photodegradation. We survey the bromine $K$ (Fig. 1a,d,g) HERFD, lead $L_3$ (Fig. 1b,e,h) HERFD and bromine VtC (Fig. 1c,f,i) spectra of the three APB's and note the following five observations.

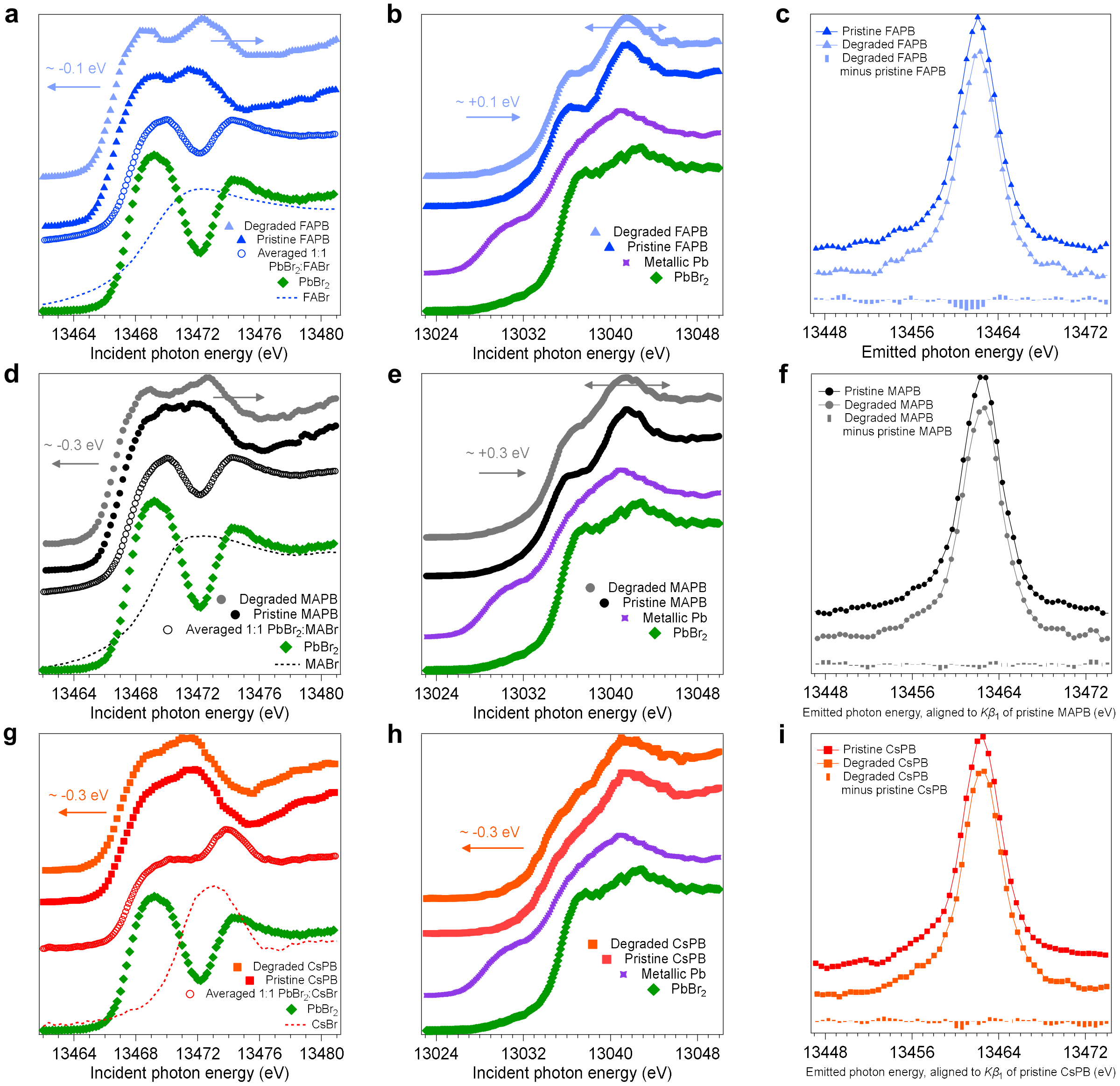

Figure 1. **Normalized bromine *K*-edge and lead $L_3$-edge X-ray absorption spectra and bromine valence-to-core X-ray emission spectra of pristine and photodegraded $APbBr_3$.** a) Bromine *K*-edge (Br *K*) High Energy Resolution Fluorescence Detected X-ray Absorption Spectroscopy (HERFD-XAS) measurements, b) lead $L_3$-edge (Pb $L_3$) HERFD-XAS measurements, and c) bromine valence-to-core (VtC) X-ray emission spectroscopy (XES) measurements of pristine and degraded FAPB. d) Br *K* HERFD-XAS measurements, e) Pb $L_3$ HERFD-XAS measurements, and f) bromine VtC XES measurements of pristine and degraded MAPB. g) Br *K* HERFD-XAS measurements, h) Pb $L_3$ HERFD-XAS measurements, and i) bromine VtC XES

measurements of pristine and degraded CsPB. The X-ray absorption spectra of the reference compounds: ABr, $PbBr_2$, 1:1 average of ABr:$PbBr_2$ for Br *K*, and metallic lead foil and $PbBr_2$ for Pb $L_3$, are plotted alongside the HERFD-XAS measurements of the pristine and photodegraded lead bromide perovskites (APB). The HERFD-XAS and bromine VtC XES spectra of the pristine compounds are similar to those presented in ref [29]. Photodegraded refers to the most degraded state accessed in our experiments. The photodegraded spectra were recorded at least ~110 minutes after the recording of the pristine spectra, from the same spot on the sample with continuous beam exposure. The dominant changes in the photodegraded HERFD spectra with respect to the pristine spectra are highlighted with arrows. The intensities of the bromine VtC XES spectra are normalized to the integrated area of bromine $K\beta_{1,3}$ (not shown here).

**Presence of photoinduced spectral and material changes**

*First*, the onsets of all photodegraded spectra are energetically shifted with respect to the pristine spectra, with the magnitudes of the shift ranging from ~0.1 to ~0.3 eV, indicating changes to Pb-Br bond ionicity upon photodegradation. In the case of the photodegraded organic APB's, the onsets of the bromine *K* and lead $L_3$ spectra shift with opposite signs and approximately the same magnitude. The photodegraded bromine *K* spectra shift down in energy, indicating that the bromide ligands have gained valence electron density, while the photodegraded lead $L_3$ spectra shift up in energy, indicating that the lead cations have lost the same quantity of valence electron density; the Pb-Br bond in the photodegraded organic APB's has become more ionic relative to the pristine case. Changes in the formal oxidation state (FOS) of lead are unlikely due to the magnitude of the shifts. The magnitude of the lead $L_3$ chemical shift for lead(II) with respect to lead(0) is +1.0 to +1.3 (± 0.2) eV, and the shift of lead(IV) with respect to lead(0) is +2.6 ± 0.2 eV [32]. Hence, lead in the photodegraded organic APB's remains lead(II); this is consistent with the conclusion we drew from the beam damage checks (Supplementary Note 2), that the photodegraded perovskites remain perovskites. We suggest that the metavalent nature of the Pb-Br bond allows it to autonomously and slightly vary its ionicity-covalency nature, or chemically flex, without changes in the FOS. The lead-halide bond has been termed metavalent for a class of chemical bonds that are neither ionic, covalent nor metallic [33]. Pure covalent bonding consists of 2.0 shared electrons and 0.0 electron transfer; pure ionic bonding consists of 1.0 relative electrons transferred and 0.0 electrons shared. In the case of the Pb-Br bond in CsPB, quantum chemical calculations show that 0.564 electrons are transferred and 0.741 electrons are shared, giving the Pb-Br bond an intermediate covalent-ionic character [33]. We expect that changes in Pb-Br bond ionicity should rehybridize the system. However, the increase in Pb-Br bond ionicity upon photodegradation may actually decrease the effects of rehybridization as the bonds are becoming less covalent. We suggest the design of metavalent chemical bonding that becomes furthermore ionic upon degradation as a general design principle for defect tolerant materials. In the case of the all-inorganic APB, we observe that the onsets of the bromine *K* and lead $L_3$ spectra of photodegraded CsPB shift down in energy with approximately the same magnitude. This shows that the Pb-Br bond ionicity has not changed upon photodegradation and that additional and comparable valence electron density is introduced onto both the bromide and lead sites upon photodegradation. Since the states in the Pb-Br sublattice are hybridized, the increase in valence electron density could arise from a lower concentration of holes or a higher concentration of electrons in the Pb-Br sublattice. Both are manifestations of self-*n*-doping due to the photoinduced formation of intrinsic defects. Observable changes to the electronic structure (bromine- and lead-projected DOS (Fig. 1g,h)) are absent given our measurement resolution, potentially due to the low concentration of defects. Defect formation energy versus Fermi energy position calculations on CsPB suggest that as the material becomes n-type, the intrinsic defect which possesses the lowest formation energy is the bromide interstitial ($Br_i$), suggesting this defect is formed upon photodegradation and is accompanied by bromide vacancy formation [34]. Our findings highlight the pivotal role of the A-cation in mediating the mechanism of photodegradation. We recall our operational definition of defect tolerance as a material property where defects have minimal effect on carrier mobility and minority-carrier lifetime. A positive correlation between electron effective mass and electrons transferred (within the lead-halide bond) in HaPs has been reported, thus we expect the increase in Pb-Br bond ionicity upon photodegradation, in the organic APB's, to have minimal effect on the electronic structure but affect the electron effective mass and hence mobility [33]. We find the presence of an inorganic A-cation leads to photoinduced self-doping via intrinsic defects that is expected to affect minority-carrier lifetime. Hence, we have found defect intolerance in all of the APB's.

*Second*, the broadening of the $e_g$ feature, and potentially the $t_{2g}$ feature also, in the lead $L_3$ spectra of the hybrid organic APBs (Fig. 1b,e) reflects photoinduced changes to the internal structure and/or cooperative tilting of the $PbBr_6$ octahedra. According to reported XAS calculations, the strongest transitions in the rising-edge (~13036 eV) and main-edge (~13041 eV) features found in the lead $L_3$ spectrum of MAPB originate from hybridized Pb *d*-Br *d* states [35]. The energy splitting between the rising- and main-edge features is dominated by the energy splitting of Pb *6d* states due to bromide ligand repulsion (octahedral ligand/crystal field splitting). While other transitions contribute to the absorption spectrum, for example hybridized Pb($d_{xy}$)-Br($d_{xy}$)-methylammonium states in the rising-edge and hybridized Pb($d_{xy}$)-methylammonium states in the main-edge, we refer to the rising-edge feature as $t_{2g}$ and the main-edge feature as $e_g$ for simplicity. Supplementary Fig. 5 shows the relative energetic ordering of $t_{2g}$ and $e_g$. Since the lead $L_3$ spectrum of MAPB is dominated by transitions to hybridized Pb *d*-Br *d* states, the octahedra in all APB are similar and all of the $L_3$ spectra contain two dominant features (Fig. 1b,e,h), we assume the $L_3$ spectra of FAPB and CsPB are dominated by similar transitions. This enables us to isolate pristine and photodegraded A-cation influence on the $PbBr_6$ octahedra. We observe that the lead $L_3$ spectra of the photodegraded organic APB's appear as shifted and broadened versions of the pristine spectra (Supplementary Fig. 6c,f). By aligning the onsets of the pristine and photodegraded spectra, we observe a decrease of the crystal field splitting in photodegraded MAPB and the broadening of $e_g$ in the photodegraded organic APB's. The $e_g$ states, which are dominated by $d_z^2$ states sigma-oriented along the Pb-Br bonds, are highly sensitive to Pb-Br bond length while the $t_{2g}$ states (comprised of $d_{xy}$, $d_{xz}$, $d_{yz}$ orbitals oriented between the Pb-Br bonds) are weakly sensitive [35]. A decrease in the crystal field splitting reflects an increase in the average Pb-Br bond length; the $PbBr_6$ octahedra are clearly distorted by photodegradation. Broadening of $e_g$ indicates the Pb-Br bond length distribution has become more disordered upon photodegradation; this points towards the formation of bromide vacancies (accompanied, presumably, by the formation of bromide interstitials and/or mobile ions). Furthermore, we observe that the number of dominant spectral features (two) remains unchanged upon photodegradation, which motivates the use of curve fitting to extract broadening trends. Ideally, curve fitting should be combined with XAS calculations. However, given current challenges with accurate modeling of defective HaP crystal structures and the dominance of only two key features, $t_{2g}$ and $e_g$, in the lead $L_3$ spectra of the APB's, even semi-quantitative trends extracted from curve fitting are valuable for guiding future computational work. We perform systematically constrained curve fitting as described in Supplementary Note 3 and shown in Supplementary Fig. 7, and find four trends. First, for the pristine APB's, the fitted FWHM of $t_{2g}$ increases, going from FAPB (5.03 eV) → MAPB (5.18 eV) → CsPB (5.74 eV). Second, for the pristine APB's, the fitted FWHM of $e_g$ increases, going from FAPB (4.04 eV) → MAPB (4.29 eV) → CsPB (5.07 eV). Third, the FWHM of $t_{2g}$ decreases when the organic APB's are degraded (FAPB: 5.03 eV → 4.88 eV, MAPB: 5.18 eV → 5.07 eV). Fourth, the FWHM of $e_g$ increases when the organic APB's are degraded (FAPB: 4.04 eV → 4.33 eV, MAPB: 4.29 eV → 4.55 eV). The energy uncertainty of the fitted FWHM is 0.04 eV. Trends one and two show that the broadening of $t_{2g}$ and $e_g$ jointly increase(decrease) as octahedral tilting increases(decreases). This can be rationalized with atomic orbital overlap; increased octahedral tilt results in more disorder in the overlap of Pb *d*-Br *d* orbitals and leads to energetic broadening. Trends three and four reveal that upon photodegradation of the organic APB, the broadening of $t_{2g}$ and $e_g$ are inversely correlated. We use this observation to suggest that the degree of octahedral tilting is unaffected by the photoinduced formation of bromide vacancies. This leads to the deduction that rehybridization of the bromine- and lead-projected DOS is absent upon photodegradation, consistent with our earlier finding that photodegradation leads to an increase in Pb-Br bond ionicity as the degree of cooperative octahedral tilting increases when Pb-Br ionicity decreases [29]. Density functional theory calculations show the degree of octahedral tilting to have a strong influence on the bandgap of MAPB, hence our finding of negligible change in octahedral tilting indicates the tolerance of band-edge optical absorption and emission to photodegradation, and is consistent with our visual observation of the XEOL color and intensity being roughly constant during photodegradation [35]. Compression (four longer B-X bonds and two shorter B-X bonds) or elongation (opposite case of compression) of the octahedron in a general $ABX_3$ perovskite will broaden both $t_{2g}$ and $e_g$, with the largest broadening/splitting in the $e_g$ level [36]. Trends three and four show that the magnitude change in FWHM is greatest for $e_g$, consistent with this rule of thumb. However, the broadening of $t_{2g}$ has decreased upon photodegradation; this could be due to a loss of hybridization between Pb($d_{xy}$)-Br($d_{xy}$) states and methylammonium states and reliable calculations are needed to gain further insight. Taken together, the observation of a decrease of the crystal field splitting and increase in the broadening of the $e_g$ level upon photodegradation indicates that the $PbBr_6$ octahedra in photodegraded MAPB and FAPB are distorted. Photoinduced, time-averaged inversion symmetry breaking could result in the emergence of exotic effects in the electronic structure (e.g. Rashba effect) during optoelectronic device operation, which we discuss later [37]. Recalling our operational definition of defect tolerance as a property where defects have minimal effect on mobility and minority-carrier lifetime, we find evidence for photoinduced distortion of the $PbBr_6$ octahedra due to bromide vacancy formation that may not perturb the electronic structure and modify carrier mobility; this is evidence for defect tolerance.

*Third*, the energetic widening of the σ-π energy separation in the bromine *K* main-edge spectra of the photodegraded organic APB's indicates that the nature and/or strength of A-Br H-bonding has changed upon photodegradation. Through the use of XAS calculations, we have previously reported the existence of two Br *4p* distributions in the bromine *K* spectra, where the states at lower energy have σ character arising from Pb-Br bonding and the states at higher energy have π character and are influenced by A-Br bonding [29]. We align the bromine *K* spectra of the photodegraded organic APB's at the onset (Supplementary Fig. 6a,d) and at the high energy side (Supplementary Fig. 6b,e) of the main-edge of the bromine *K* spectra of the pristine organic APB's and note three observations. First, the distribution of states at the onset (below ~13468.5 eV), associated with the hybridized Pb *6p*-Br *4p* states near the CBM, is unaffected by photodegradation in both of the organic APB's. This indicates that the Br *4p* PDOS profile, and likely the total density of states (TDOS), near the CBM, within measurement resolution limits, is unaffected by photodegradation and provides evidence for defect tolerance. Second, a loss of Br *4p* σ states upon photodegradation is observed in both organic APB's. This is evidence for the formation of bromide vacancies; the σ states originate from (anti-bonding) Pb-Br states [29]. Third, the distribution of states at the high energy side of the main-edge is unaffected by photodegradation. This confirms the widening of the σ-π energy separation since the distribution of states at the onset and high energy side of the main-edge are unaffected by photodegradation. As for CsPB, which lacks A-Br H-bonding, aside from the shift of the onset of the photodegraded spectrum with respect to the pristine spectrum as noted earlier, no change, within the resolution limits of our measurement, is observed. We have previously found a positive correlation between A-Br H-bonding strength and magnitude of σ-π energy separation; the widening of σ-π indicates that the nature and/or strength of H-bonding has changed upon photodegradation [29]. To investigate the mechanism of organic A-cation photodegradation, we examine the bromine *K* difference spectra of MAPB (Fig. 2a), generated by subtracting the spectra for spectrometer runs 3 to 13 from the pristine spectrum. We observe that the difference spectra show the evolution of spectral features in three energy regions: < 13468.5 eV, ~13468.5 to ~13472.5 eV, and > 13472.5 eV, defined by the zero-crossing points of the difference spectra, which we refer to as regions A, B and C, respectively. Region A is predominantly sensitive to shifts of the onset, region B is predominantly sensitive to changes of the σ distribution and region C is predominantly sensitive to changes of the π distribution. Based on the temporal evolution of the relative intensities of the features in regions A, B and C, we group the spectrometer runs into three temporal regimes that we label as regimes I, II and III, respectively. Regime I, comprising runs 3 and 4, shows negligible changes in energy regions A to C. Regime II, encompassing runs 5 to 10, shows positive intensity in region A (onset shifted to lower energy) and negative intensity in region B (loss of σ states). Regime III, encompassing runs 11 to 13, shows noticeably higher positive intensity in region A (larger onset shift relative to regime II), noticeably higher negative intensity in region B (increased loss of σ states relative to regime II) and new positive intensity in region C (π states shifted to higher energy). We generate representative spectra from regimes II and III and plot them in Fig. 2b, along with the labels for energy regions A to C. To validate the two regimes of onset shifts observed in region A, we perform a similar procedure with the lead $L_3$ difference spectra (Fig. 2c,d), described in Supplementary Note 4. We observe that regimes I to III for the bromine *K* and lead $L_3$ difference spectra encompass roughly the same ranges of spectrometer runs. This close match, facilitated by controlled photodegradation conditions and intentional optimization of spectrometer runtimes to be comparable for the two absorption edges, enables us to follow concurrent changes in Pb-Br bonding, A-Br bonding, $PbBr_6$ structure. In regime II, the formation of bromide vacancies (loss of σ states, Fig. 2a) is accompanied by an increase in Pb-Br bond ionicity (onset shifts, Fig. 2a and 2c) and time-averaged internal structural distortion of the $PbBr_6$ octahedra (decrease of crystal field splitting, broadening of $e_g$ Fig. 2c). The Br *4p* π states are unaffected in regime II, suggesting the N-H…Br hydrogen-bonding accommodates the increase in Pb-Br bond ionicity; we have previously found a positive correlation between H-bonding strength and Pb-Br bond ionicity in the pristine APB's, revealing that the A-Br and Pb-Br bonding are interlinked [29]. Since photodegraded CsPB contains self-*n*-doping defects, likely due to the formation of bromide vacancies, and no evidence for self-doping exists for the photodegraded organic APB's, in spite of the observed formation of bromide vacancies, we deduce that the existence of A-Br hydrogen-bonding mitigates the formation of photoinduced intrinsic defects. We have suggested that the metavalent nature of the Pb-Br bond enables it to chemically flex and have found that the existence of A-Br H-bonding facilitates the chemical flexing of the Pb-Br bond. After over a century of research, the nature of the hydrogen bond is not fully understood yet, though the hydrogen bond has been reported to feature broad transition regions that merge continuously with the covalent bond, van der Waals interaction, ionic interaction, etc [38,39]. Based on this known phenomenon, and our finding that the Pb-Br bond can chemically flex in the presence of A-Br H-bonding, we suggest that autonomous chemical bond flexibility in all of the chemical bonds in a material system is a general design principle for defect tolerant materials. In regime III, a further increase in the concentration of bromide vacancies (Fig. 2a) is accompanied by a further increase in Pb-Br bond ionicity, as observed from the onsets of both the bromine (Fig. 2a) and lead (Fig. 2c) spectra, further distortion of the $PbBr_6$ octahedra and a shift of the Br *4p* π states to higher energy. This shift of the π states occurs in tandem with the abrupt increase in Pb-Br bond ionicity. We have probed two distinct regimes of photodegradation of the organic APB's here; more may exist. Nevertheless, the photodegraded organic APB's remain perovskites, according to our beam damage checks (Supplementary Note 2); organic A-cations in the probed sample region have not diffused away en masse and remain, in some chemical form. The N-H…Br bond may transition from a conventional, electrostatic H-bond to a stronger, hydrogen-mediated chemical bond or cleave, hence forming $MA^+$/$FA^+$ interstitials that still interact strongly with the Pb-Br sublattice. Irrespective of the specific mechanism(s) of A-cation H-bond photodegradation in regime III, the Pb-Br bonding is capable of chemically flexing further. We recall our operational definition of defect tolerance as a material property where defects have minimal effect on carrier mobility and minority-carrier lifetime. We find the distribution of states at the onset (below ~13468.5 eV), associated with the hybridized Pb *6p*-Br *4p* states near the CBM, is unaffected by the chemical and structural effects of photodegradation in both of the organic APB's and hence find evidence for defect tolerance. Widening of the σ-π energy separation is expected to have negligible impact on electron dynamics at the CBM since we have previously found the σ-π energy separation to be relevant for hot electron cooling/dynamics, and may even decrease hot electron cooling rate [29]. The loss of Br *4p* σ states is expected to reduce the TDOS near the CBM, which may or may not affect optoelectronic functionality depending on the magnitude of excess carrier (photo)generation.

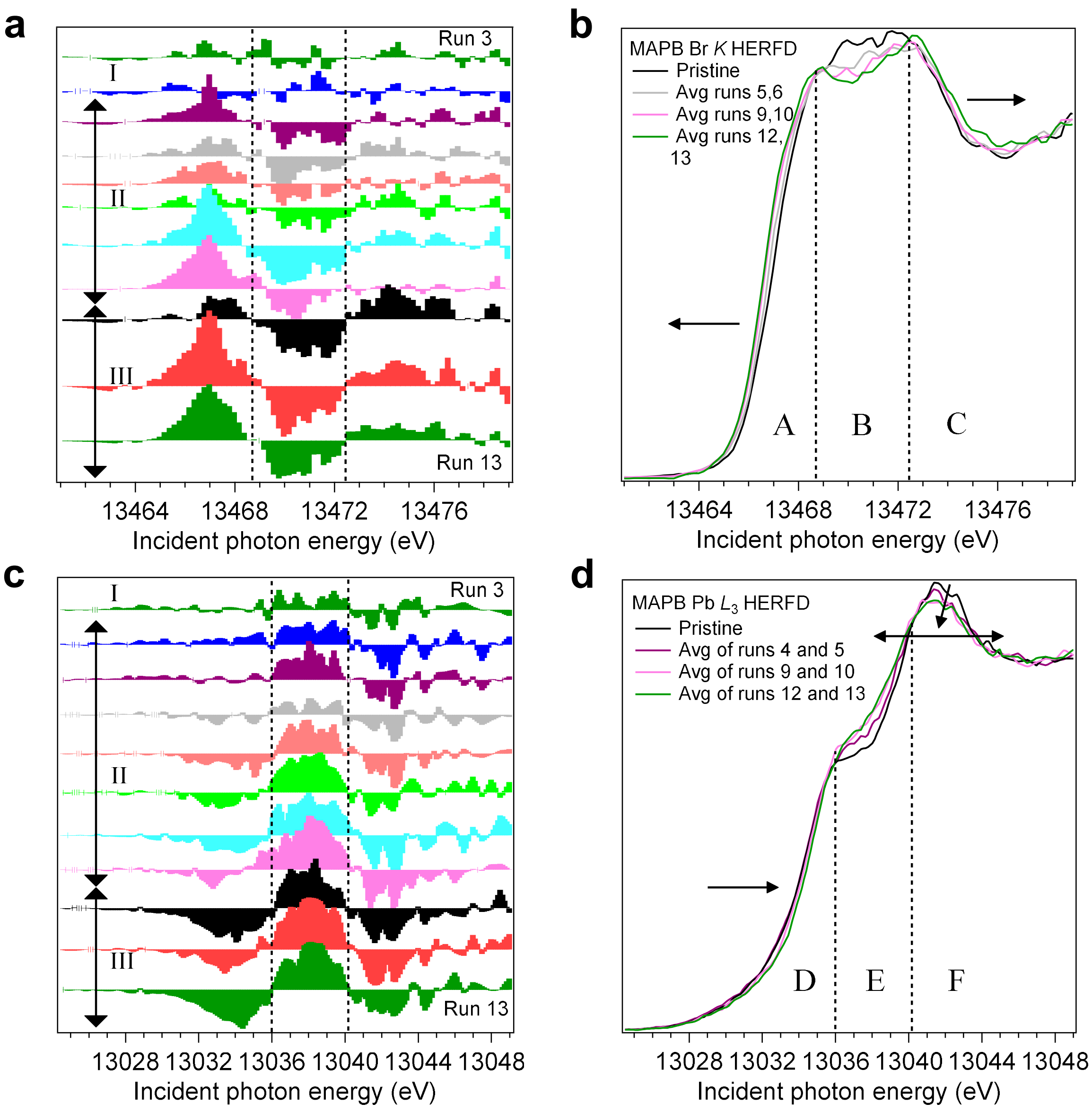

Figure 2**. Normalized and difference bromine *K*-edge and lead $L_3$-edge X-ray absorption spectra recorded from continuously photodegrading MAPB.** a) Bromine *K*-edge difference spectra generated by subtracting the spectra for spectrometer runs 3 to 13 from the pristine spectrum. The energy window is separated into three regions: A, B, C, at the zero-crossing points of the difference spectra. The spectrometer runs are grouped into three temporal regimes: I, II, III, based on similarities in the difference spectra. b) Representative spectra from regimes II and III, plotted against the pristine spectrum. c) Lead $L_3$-edge difference spectra generated by subtracting the spectra for spectrometer runs 3 to 13 from the pristine spectrum. The energy window is separated into three regions: D, E, F, at the zero-crossing points of the difference spectra. The spectrometer runs are grouped into three temporal regimes: I, II, III, based on similarities in the difference spectra. d) Representative spectra from regimes II and III, plotted against the pristine spectrum.

**Absence of photoinduced spectral and material changes**

*Fourth*, the measured distributions of states at the onset/pre-edge region (~13462 to ~13467 eV for bromine *K*-edge, ~13026 to ~13034 eV for lead $L_3$) in all of the APB are unaffected by photodegradation. Energetic alignment of the onsets of the photodegraded and pristine spectra confirms this point (Supplementary Fig. 6a,c,d,f,g,h). Hence, we find the absence of major changes to the distributions of Br *4p* and Pb *6p* states near the CBM upon photodegradation and confirm our earlier deduction, based on Pb-Br bond ionicity and $t_{2g}$ and $e_g$ broadening evidence, that the formation of bromide vacancies has minimal effect on rehybridization of lead-bromide states. Earlier, we found an autonomous increase to the Pb-Br bond ionicity in response to photodegradation of the organic APB. This change is expected to shift the relative weights of lead and bromide contributions to the CBM (and VBM) and consequently modify the electron (and hole) effective mass, and could be related to the reported positive correlation between electron effective mass and electrons transferred in HaPs [33]. Recalling our operational definition of defect tolerance as a property where defects have minimal effect on mobility and minority-carrier lifetime, we find the presence of photoinduced bromide vacancies to have minimal effect on the cooperative octahedral tilting and the bromine- and lead-PDOS near the CBM. This is evidence for defect tolerance in the electronic structure to structural and compositional defects. However, changes in carrier effective masses are expected, which necessitates a precise definition of what minimal effect is.

*Fifth*, the pristine and photodegraded VtC spectra for each APB (Fig. 1c,f,i) are the same, given our ~1.5 eV spectrometer resolution. This observation indicates that the distribution of Br *4p* states close to the VBM is insensitive to the formation of bromide vacancies. Recalling our operational definition of defect tolerance as a property where defects have minimal effect on mobility and minority-carrier lifetime, we find further evidence for defect tolerance of the electronic structure to structural and compositional defects. We note that while the measured distributions of bromine *4p* states near the VBM appear unchanged upon photodegradation, suggesting the hole effective mass is unaffected by photodegradation, a change is expected due to the change in Pb-Br bond ionicity as mentioned for observation four.

## DISCUSSION

Our findings highlight the pivotal role of the A-cation in mediating the influence of intrinsic defects; the presence of A-Br hydrogen-bonding in the organic APB mitigates bromide vacancy-induced self-doping by enabling the Pb-Br bonding to chemically flex. Quantum chemical calculations of the Cs-Br bond in CsPB show 0.983 electrons transferred and 0.094 electrons shared [33]. We observe that the Cs-Br bond in CsPB is nearly purely ionic and likely does not possess chemical bond flexibility. The type of A-cation influences the rate of photodegradation also. Given comparable X-ray doses, substantially more bromide vacancies are formed in the organic APB's; the decrease of the σ states is noticeable in the spectra of the photodegraded organic APB's but not in the spectrum of photodegraded CsPB.

We discuss the implications of our work for the definition of defect tolerance, the apparent contradiction between defect tolerance and organic HaP device ambient instability, the suitability of HaPs as a > 20 year lifetime solar energy conversion technology, all beam-based measurements of HaPs, DFT-based computation and general design principles for defect tolerance in materials. Throughout this work, our operational definition of defect tolerance is a qualitative statement: a material property where the defects that do form have minimal effect on mobility (μ) and minority-carrier lifetime (τ) [3]. We find, for the organic APB's, that the measured distributions of key components of the electronic structure - the bromine *4p* states near the VBM, lead *6p* states near the CBM and bromine *4p* states near the CBM - are insensitive to (A-cation)-Br and Pb-Br bonding changes and time-averaged distortion of the $PbBr_6$ octahedra due to vacancy formation. This is evidence for tolerance of the electronic structure to structural and compositional defects; however, (potentially small) changes to the effective masses and hence carrier mobility are expected due to the small change in Pb-Br bond ionicity. Our work highlights the need to precisely define what "minimal effect" is. For CsPB, self-*n*-doping of the material upon photodegradation may result in a (potentially small) change to the minority-carrier lifetime.

The purported defect tolerance of HaPs seemingly contradicts the numerous reports on the ambient chemical instability of HaPs. We offer a resolution based on A-cation-influenced differences in the interaction of bromide interstitial defects with the Pb-Br sublattice that may be generalizable to all HaPs. This explanation is relevant for another ongoing puzzle, that in spite of the purported superior ambient stability of cesium-containing HaPs, the highest performing solar cells are based only on organic/$FA^+$-containing HaP [8,40]. Self-healing in the same APB's investigated here has been reported; the discussed self-healing mechanisms involved decomposition into and reformation from ABr and $PbBr_2$ [41]. We suggest that self-healing is possible also in the damage regime we have explored here, which does not involve decomposition into the binary precursor compounds, and is facilitated by the combined chemical bond flexibility of the (A-cation)-Br and Pb-Br bonds in the organic APBs. Upon photodegradation, the bromide interstitial defects in the organic APB's likely diffuse away from the damaged region as they do not appear to interact strongly with the Pb-Br sublattice. In contrast, the bromide interstitials in photodegraded CsPB do interact strongly with (and dope) the Pb-Br sublattice. The expected consequences of weak sublattice-defect interaction are electrical/electronic insensitivity to mobile ionic defects, relatively higher ion diffusion coefficients and relatively higher self-healing rates (the primary ion-sublattice interaction mechanism is vacancy elimination and not doping). The expected consequences of strong sublattice-defect interaction are the converse. Formaminium is reported to best self-heal the bulk material while cesium protects the surface best [42]. This is consistent with our deduction. The all-inorganic, cesium-containing HaPs appear to possess higher ambient stability as their mobile ionic defects likely possess lower diffusion coefficients; relative to the organic-containing HaPs, it takes longer for the mobile ions to chemically react with the contacts in a device, for example, rendering them unavailable to self-heal the damaged perovskite. Insensitivity to defects in general is a key consideration for a solar absorber, hence organic-based HaPs are preferred. Our work here shows that the electronic structure of organic APB's is (partially) tolerant to photodegradation. Cahen *et al.* [11] ask "are defects in lead-halide perovskites healed, tolerated, or both?" and we answer "both" based on work reported in the literature and our findings here. If the mobile halide interstitial defects in organic HaPs remain available to self-heal the perovskite, we see no intrinsic reasons for why organic-based HaPs cannot form the basis of a > 20 year lifetime solar energy conversion technology. In fact, the photogeneration of mobile ions serves as an energy storage mechanism (photo-rechargeable perovskite batteries have been reported) and an excess energy dissipation mechanism (hence prolonging the lifetime of the absorber) [43]. The ions could be trapped within a closed, chemically inert environment, for example with chemically inert contacts/electrodes, and/or isolated from grain boundaries and surfaces with additives (an approach used in the highest-efficiency solar cells) [8]. We note that the use of HaPs as a solar absorber material places particular demands on the design of the device interfaces, as the presence and elimination of mobile ions affects energy level alignment, charge extraction rates, etc.; from a device/technology standpoint, HaPs may be "all about the interfaces" [44]. The use of mixed A-cation HaP films was originally justified based on entropic stabilization of photoactive perovskite phases [45]. We suggest a complementary, practical synergistic reason, that of combining the best (from an application standpoint) properties of weak and strong ion/defect-sublattice interaction.

Measurements of fundamental (photo)physical properties of HaPs are typically performed with beam-based measurements, using nonionizing or ionizing light, electrons, etc., and the sensitivity of HaPs to beams is known [22]. The conventional approach to assessing beam damage is to monitor the material for decomposition into its constituents, using for example core level photoelectron spectroscopy (PES) or photoluminescence to search for chemical changes to the perovskite [23,41]. Our work reveals a photodegradation regime where the perovskite has not decomposed into its binary precursor compounds, is intact but is defective due to halide vacancies, and time-averaged inversion symmetry breaking is present in the $PbBr_6$ octahedra of MAPB and FAPB. Using the same technique (angle-resolved photoelectron spectroscopy (ARPES)) applied to the same material (MAPB), both in single crystal form, different groups have concluded that Rashba effects are giant or negligible [46,47]. We suggest that the durations of data acquisition time relative to photodegradation and self-healing time may explain the discrepancy. High-sensitivity ultraviolet photoelectron spectroscopy (HS-UPS) was used to measure the distribution of gap states in the bandgap of single crystal MAPB [48]. If the sample was photodegraded (but remains a perovskite) during measurement, the expected increase in Pb-Br bond ionicity would shift the distribution(s) of lead *6s*/*6p* states towards the Fermi energy with respect to the bromine *4p* states; the gap state distribution could include contributions from "intrinsic" lead states. Our work motivates the (re)evaluation of photophysical properties of HaPs supported with quantitative measurements of the photodegradation state. The use of optical emission energy from photo-/electro-/etc. luminescence measurements to monitor the degradation state of HaPs, in the regime where the degraded perovskite remains a perovskite, may not be fruitful; we found earlier that in spite of bromide vacancy-induced distortions of the $PbBr_6$ octahedra, the degree of cooperative octahedral tilting and hence bandgap value, remains unperturbed, consistent with our visual observation that the XEOL color and intensity remained roughly unchanged during the course of photodegradation.

Current computational approaches to model the crystal and electronic structures of pristine and defective HaPs rely heavily on DFT, yet material-independent (i.e. single exchange-correlation functional), accurate calculations of the magnitude of the bandgaps of pristine HaPs still presents significant challenges [49]. The degree of electron transfer in the lead-halide bond strongly influences the magnitude of the bandgap, as has been exploited by varying the halide species in $APbX_3$ (where X = iodide, bromide, chloride) [33]. Our work reveals that small changes in lead-halide bond ionicity, or alternatively the electrons transferred, accompany the formation of halide vacancies. To model the crystal and electronic structures of defective HaPs accurately with DFT, the first prerequisite appears to be accurate, material-independent and tuning parameter-free calculation of HaP bandgap magnitudes.

Where the flexibility to design a new compound is available, five design principles for defect tolerance have been reported: (i) formation of (energetically) shallow defects, (ii) enhanced screening, (iii) low carrier mass, (iv) kinetics and temperature control and (v) benign defect complexes [1]. The self-*n*-doping of CsPB upon photodegradation suggests the bromide interstitial is an energetically shallow defect. The dielectric constant of HaPs are relatively high and the increase of the Pb-Br bond ionicity in the organic

APB's upon photodegradation suggests the ability of the perovskite to screen charged defects is enhanced upon photodegradation. The hole and electron masses of HaPs have been calculated to be small, and our measurements suggest photoinduced changes to the effective mass (due to shifts of the Pb-Br bond ionicity) could be small [33]. We have found evidence that bromide interstitials are benign in the organic APB's. Lead halide perovskites fulfill all of the listed criteria for defect tolerance. Through our work here, we add two more design principles to the list. The first is chemical bond flexibility, in terms of autonomous reallocation of electrons transferred and shared, in all of the bonds in a material system. This implies the avoidance of materials containing purely or nearly-purely ionic or covalent, chemically inflexible bonding in sustainability-minded applications. The second is the contribution of states to the VBM and CBM by metavalent or semi-ionic bonds, whose bonds become furthermore ionic upon the formation of structural/compositional defects and hence mitigate re-hybridization of the electronic structure.

As the entire family of perovskites exhibits defect-induced polymorphism [36], we foresee the approach we have demonstrated being applied to all perovskites including oxide and two-dimensional/etc. halide perovskites, thus unraveling unexplored structure-property relationships in these materials and furthering the development of optoelectronic devices.

## Methods

**APB crystal growth.** The MAPB and FAPB single crystals were solution-grown using methods reported by Dr. Pabitra Nayak previously [50,51]. The CsPB single crystals were solution-grown using a method reported by Dirin *et al.* [26]. All single crystals were grown in an ambient air environment, then transferred into vials of chlorobenzene for storage and preservation. We have previously performed X-ray diffraction measurements of our crystals and found the crystallographic parameters to be consistent with those reported in the literature [29].

**Hard X-ray absorption and emission spectroscopy.** The spectroscopic measurements were performed (on APB crystals from the same batch characterized with XRD) at beamline P64 of the synchrotron facility PETRA III [52]. Incident photon energy calibration at the Pb $L_3$ edge was performed with metallic lead foil. The photon flux incident on the sample at ~13 keV was ~3 x $10^{11}$ photons $s^{-1}$. The X-ray spot size, measured with the X-ray eye, is ~220 µm x 100 µm. Total FY-XAS measurements were recorded with a passivated implanted planar silicon (PIPS) detector (Canberra). Resonant XES maps were recorded using a von Hamos-type hard X-ray crystal spectrometer mounted in a Bragg scattering configuration. The spectrometer featured 8 crystal analyzers; the third order reflection of the Si(220) crystal analyzers was used for Pb $L\alpha_{1,2}$ measurements and the fourth order reflection was used for Br $K\beta_{1,2,3}$ measurements[53]. Energy calibration of the two-dimensional detector images was performed with custom Python-based software written for P64 by Dr. Aleksandr Kalinko. The Br *K* maps were energy-calibrated using several elastic lines which span the spectrometer energy window, while the Pb $L_3$ maps were energy-calibrated by setting the Pb $L\alpha_{1,2}$ emission energies to standard values of 10551 and 10449 eV, respectively. Bromine *K* (~13.47 keV) and lead $L_3$ (~13.04 keV) HERFD-XAS spectra were generated as averaged intensities of the slice cut from RXES maps. The energetic width of the slice on the emitted energy axis (HERFD linewidth) was 1.6 eV and the cut was done through the RXES maximum. We note that as the organic APB's photodegrade, the energetic position of the Br *K* RXES maximum fluctuates by up to ~0.15 eV from the initial value. The HERFD linewidth was selected based on a balance between energy resolution and signal-to-noise. It was shown previously that for material systems with relatively delocalized states, such a RXES map cut is a good higher-resolution approximation to the conventional XAS spectrum [54]. The off-resonant, higher incident energy half of the RXES map was used to generate the X-ray emission spectrum. The energy resolution of XES is determined by the ~0.3 eV photon bandwidth of the Si(311) monochromator and the ~1.5 eV FWHM Gaussian broadening of the spectrometer.

Single crystals were prepared for measurements by first extracting them from chlorobenzene, blowdrying with compressed air/nitrogen in an ambient air environment, then mounting them into a Linkam T95 heating/cooling stage (~ 5 minutes total). All measurements on the APB single crystals were performed inside the Linkam stage, purged and filled with dry nitrogen, at room temperature. Halide perovskite compounds are known to degrade upon irradiation with energetic beams[22,24]. We optimized the beamline, spectrometer and measurement settings to obtain at least two iterations of HERFD-XAS spectra, recorded sequentially from the same (initially fresh) spot of a single crystal sample, which do not show observable differences due to beam damage. The measurement duration of each Br *K* and Pb $L_3$ spectrometer iteration/run was optimized to be ~12 minutes long. The photon flux was monitored and tuned, if needed, to yield comparable dose per spectrometer run for the different APB compounds. The spectra of the binary precursor compounds $PbBr_2$ and MABr are consistent with previous reports [35].

**Data availability**

The data presented in this work are available from the corresponding authors on reasonable request.

**Acknowledgements**

S.M.B. and G.J.M. thank the Swedish Research Council (# 2018-05525) for financial support. H.R., G.J.M., and D.P. acknowledge the Swedish Research Council (# 2018-06465 and # 2018-04330) and the Swedish Energy Agency (P50636) for funding. D.P. thanks the Swedish Research Council for support (# 2020-00681). P.K.N. acknowledges support from the Department of Atomic Energy, Government of India, under Project Identification no. RTI 4007, Science and Engineering Research Board India core research grant (CRG/2020/003877) and Swarna Jayanti Fellowship, DST, India.

We acknowledge DESY (Hamburg, Germany), a member of the Helmholtz Association HGF, for the provision of experimental facilities. Parts of this research were carried out at PETRA III and we would like to thank Wolfgang Caliebe for assistance in using beamline P64. Beamtime was allocated for proposals I-20181028 EC and I-20190356 EC. The research leading to this result has been supported by the project CALIPSOplus under the Grant Agreement 730872 from the EU Framework Programme for Research and Innovation HORIZON 2020. The von Hamos-type hard X-ray spectrometer was realized by the group of Prof. Matthias Bauer (University of Paderborn) in the frame of projects FKZ 05K13UK1 and FKZ 05K14PP1, supported by Bundesministerium für Bildung und Forschung.

**Author contributions**

G.J.M. and S.M.B. conceived and designed the experimental study. P.K.N. grew the single crystals and performed XRD characterization. G.J.M., S.M.B., A.K. and D.P. performed X-ray spectroscopic measurements and analyzed the data. G.J.M. and S.M.B. wrote the first draft of the manuscript. All authors contributed to revisions of the manuscript. G.J.M., H.R. and S.M.B. supervised the project.

**Competing interests**

The authors declare no competing interests.

# Supplementary Information - Experimentally observed defect tolerance in the electronic structure of lead bromide perovskites

Gabriel J. Man[1,2,3*], Aleksandr Kalinko[4], Dibya Phuyal[5], Pabitra K. Nayak[6], Håkan Rensmo[1], Sergei M. Butorin[1*]

**Author addresses:**

1. Condensed Matter Physics of Energy Materials, Division of X-ray Photon Science, Department of Physics and Astronomy, Uppsala University, Box 516, Uppsala 75121, Sweden
2. GJM Scientific Consulting, Fort Lee, New Jersey 07024, USA
3. MAX IV Laboratory, Lund University, PO Box 118, Lund 22100, Sweden
4. Deutsches Elektronen-Synchrotron DESY, Notkestraße 85, Hamburg 22607, Germany
5. Division of Material and Nano Physics, Department of Applied Physics, KTH Royal Institute of Technology, Stockholm 10691, Sweden
6. Tata Institute of Fundamental Research, 36/P, Gopanpally Village, Serilingampally Mandal, Hyderabad 500046, India

* Correspondence and requests for materials addressed to gabriel.man@maxiv.lu.se and sergei.butorin@physics.uu.se.

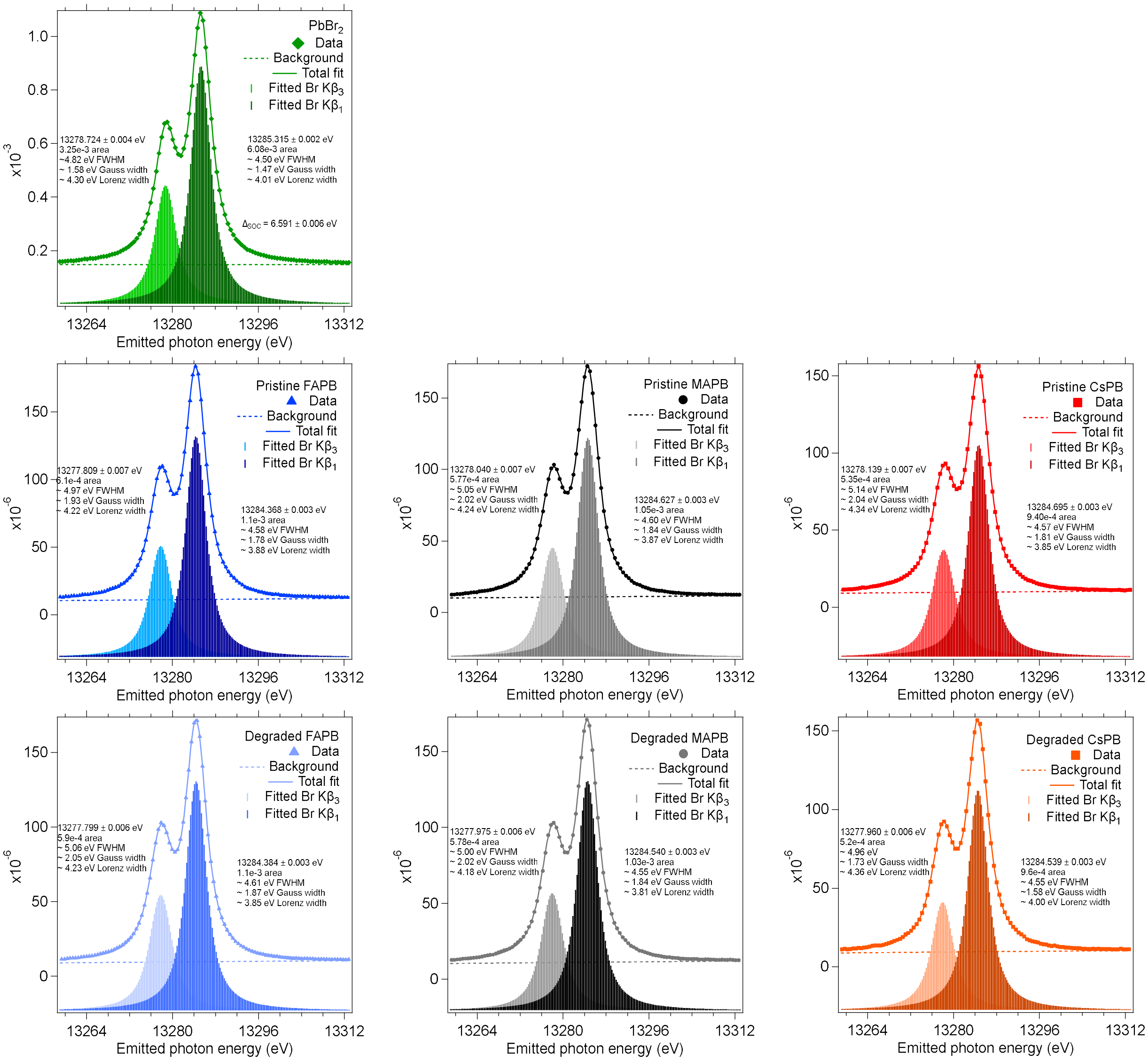

Supplementary Fig. 1. **Comparison of bromine $K\beta_{1,3}$ X-ray emission spectra of $PbBr_2$, pristine and photodegraded FAPB, MAPB and CsPB**. The curve-fitted parameters are summarized in Supplementary Table 1.

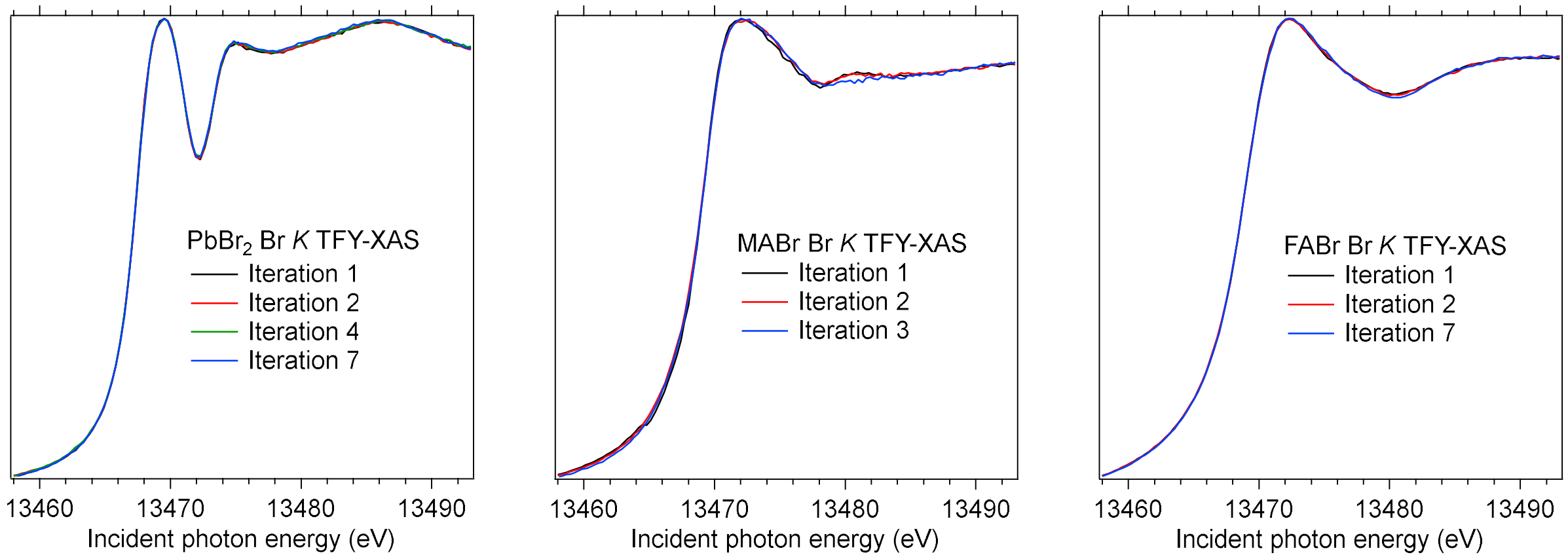

Supplementary Fig. 2. **Beam damage checks of the perovskite precursor compounds $PbBr_2$, MABr and FABr.** The checks are performed with multiple iterations of bromine *K*-edge total fluorescence yield X-ray absorption spectroscopy (TFY-XAS) measurements.

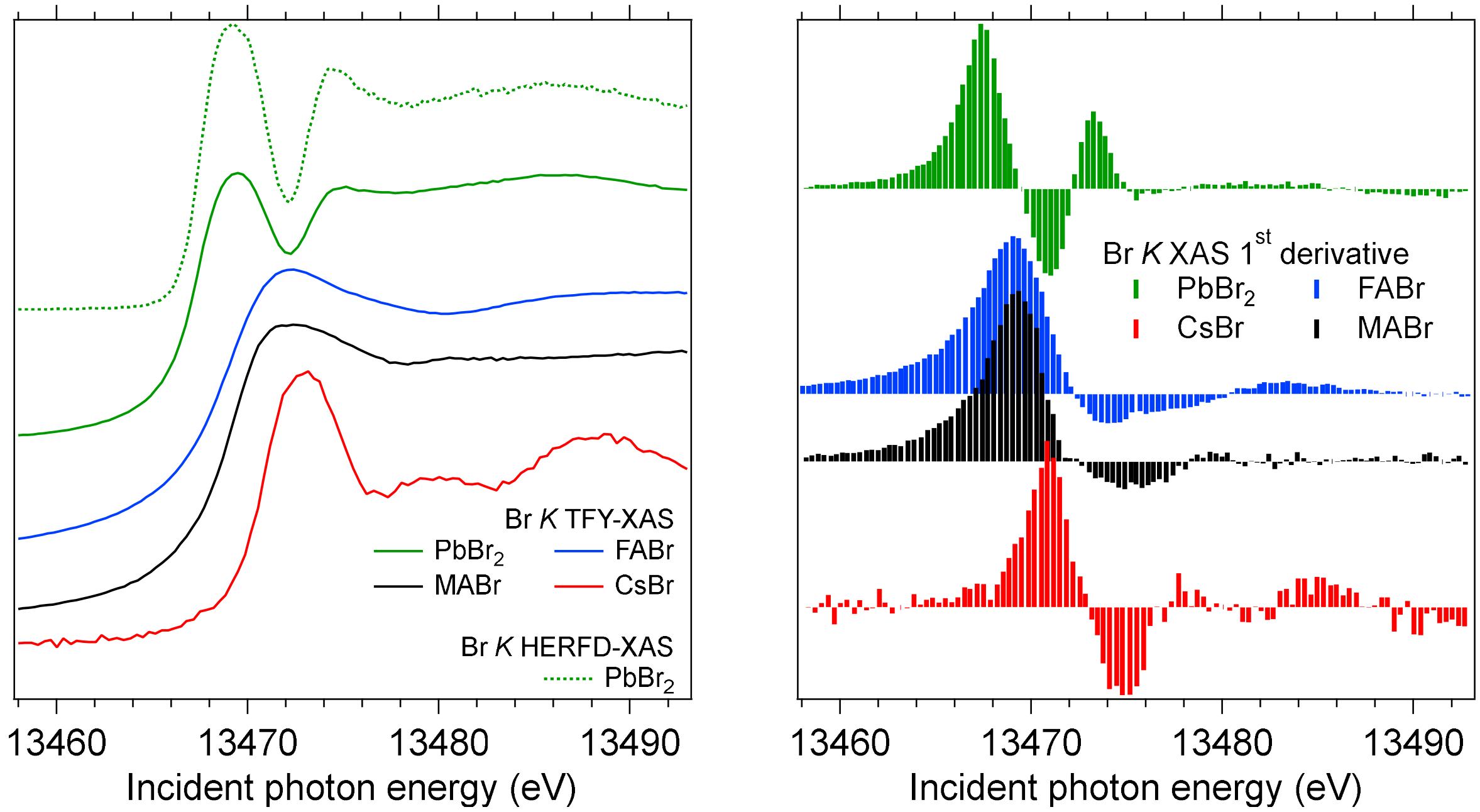

Supplementary Fig. 3. **Comparison of bromine *K*-edge X-ray absorption spectra of $PbBr_2$, FABr, MABr and CsBr. left** The comparison includes as-recorded X-ray absorption spectra (XAS) and **right** the first derivative of the XAS.

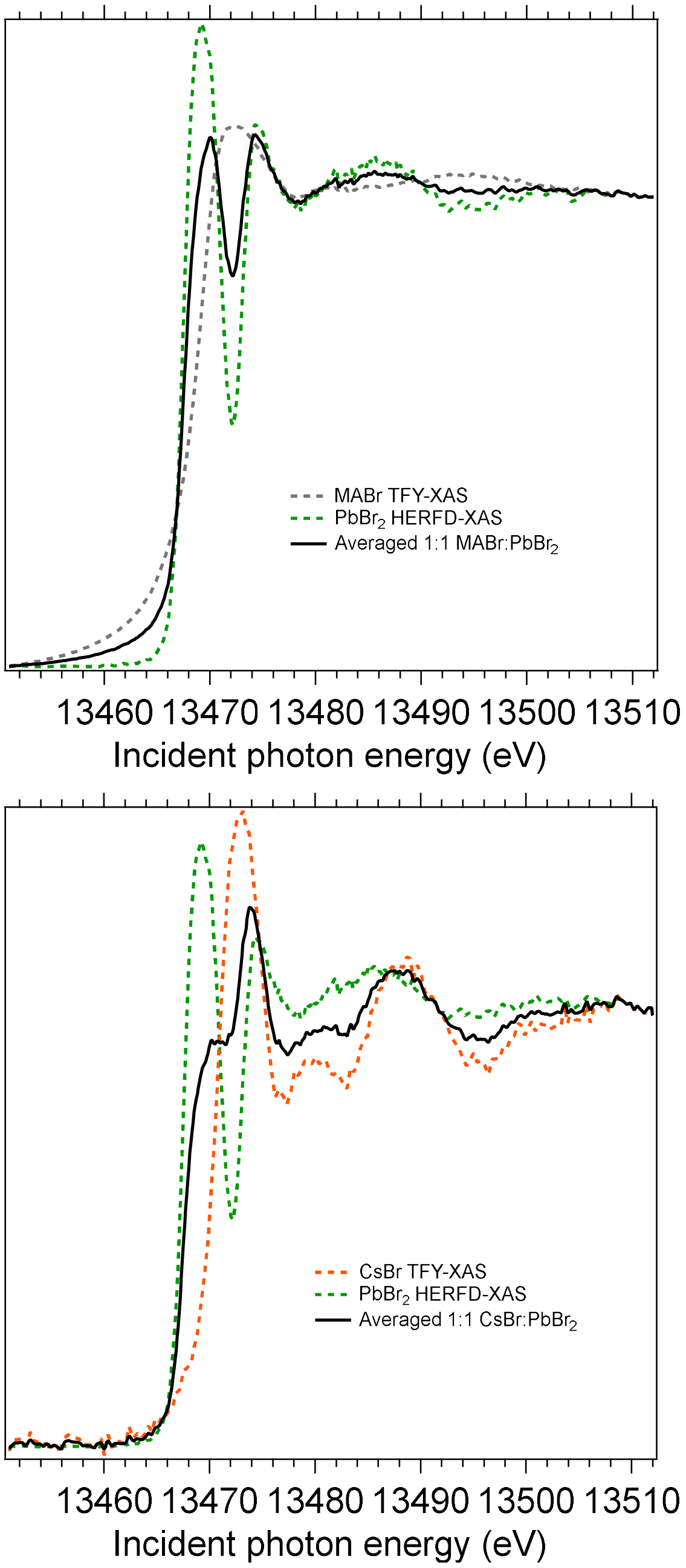

Supplementary Fig. 4. **Simulated bromine *K*-edge X-ray absorption spectra of the perovskites fully decomposed into $PbBr_2$ and ABr.** The simulated bromine *K*-edge X-ray absorption spectra (Br *K* XAS) of decomposed FAPB, MAPB and CsPB are obtained via 1:1 averaging of normalized Br *K* XAS spectra recorded from $PbBr_2$ and ABr, where A = FA/MA/Cs.

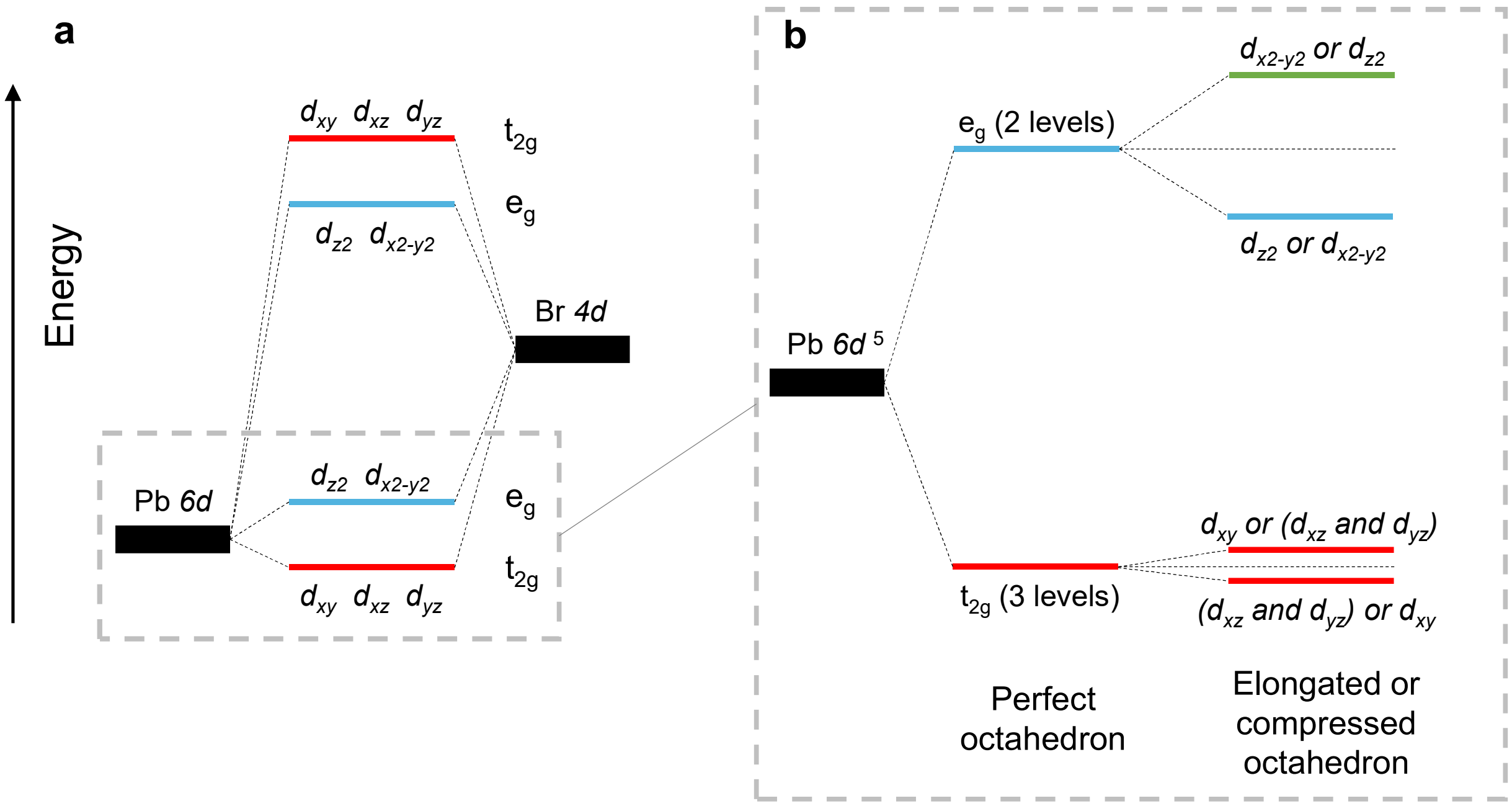

Supplementary Fig. 5. **Simplified energy level diagram showing the hybridization of lead *6d* and bromine *4d* states yielding $t_{2g}$ and $e_g$ levels.** The lead *6d* crystal field splitting diagram is inspired by references [35,36]. The position of the Fermi energy is approximately 10 eV below. The inset shows how the $t_{2g}$ and $e_g$ energy levels shift as the $PbBr_6$ octahedron is elongated or compressed.

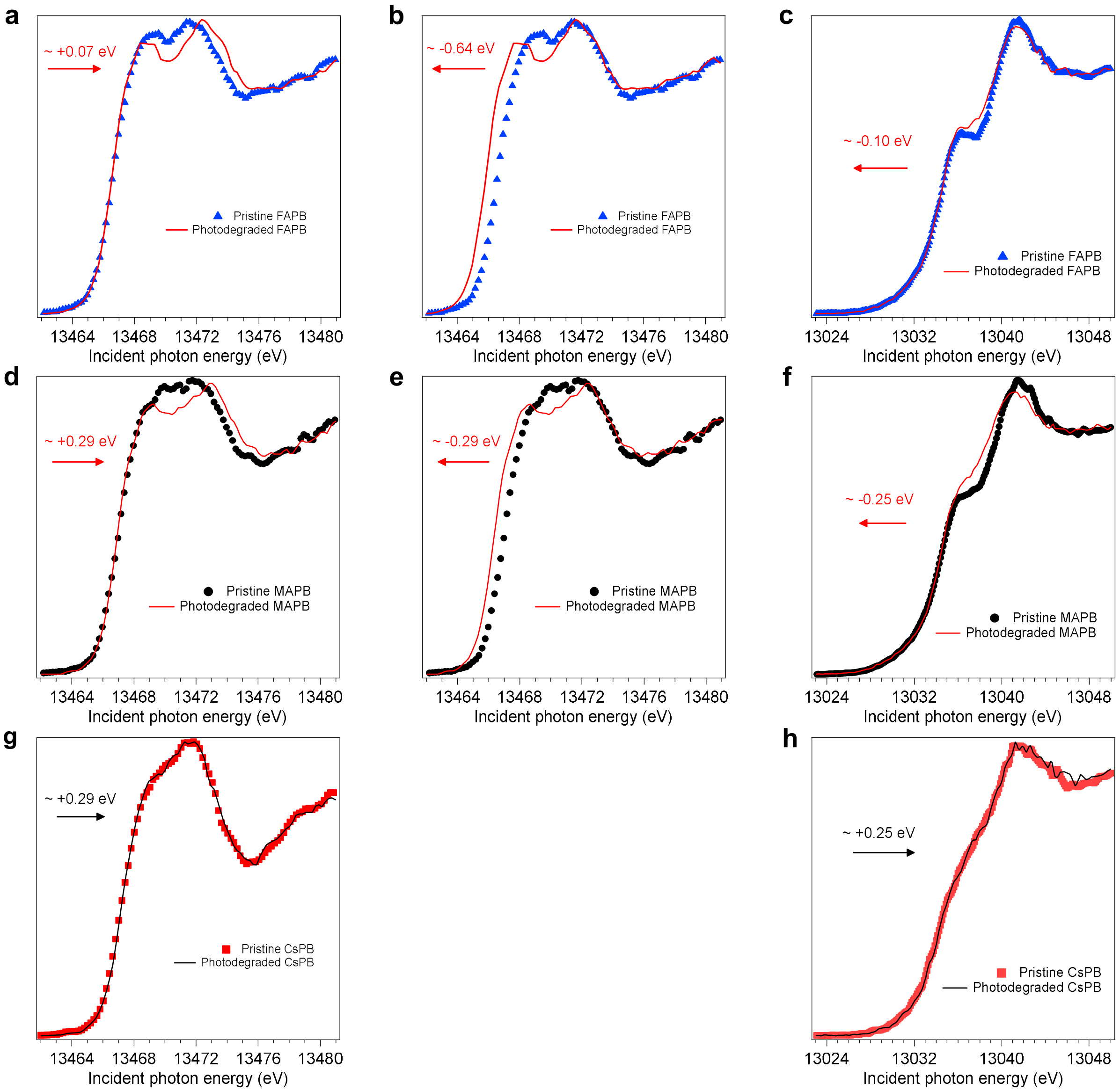

Supplementary Fig. 6. **Normalized bromine *K*-edge and lead $L_3$-edge X-ray absorption spectra of pristine versus photodegraded APB.** a) Bromine *K*-edge (Br *K*) High Energy Resolution Fluorescence Detected X-ray Absorption Spectroscopy (HERFD-XAS) measurements of pristine and photodegraded FAPB aligned at the onset, and b) aligned at the high-energy side of the main-edge. c) Lead $L_3$-edge (Pb $L_3$) HERFD-XAS measurements of pristine and photodegraded FAPB aligned at the onset. d) Br *K* HERFD-XAS measurements of pristine and photodegraded MAPB aligned at the onset, and e) aligned at the high-energy side of the main-edge. f) Pb $L_3$ HERFD-XAS measurements of pristine and photodegraded MAPB aligned at the onset. g) Br *K* HERFD-XAS measurements of pristine and photodegraded CsPB aligned at the onset, and h) Pb $L_3$ HERFD-XAS measurements of pristine and photodegraded CsPB aligned at the onset.

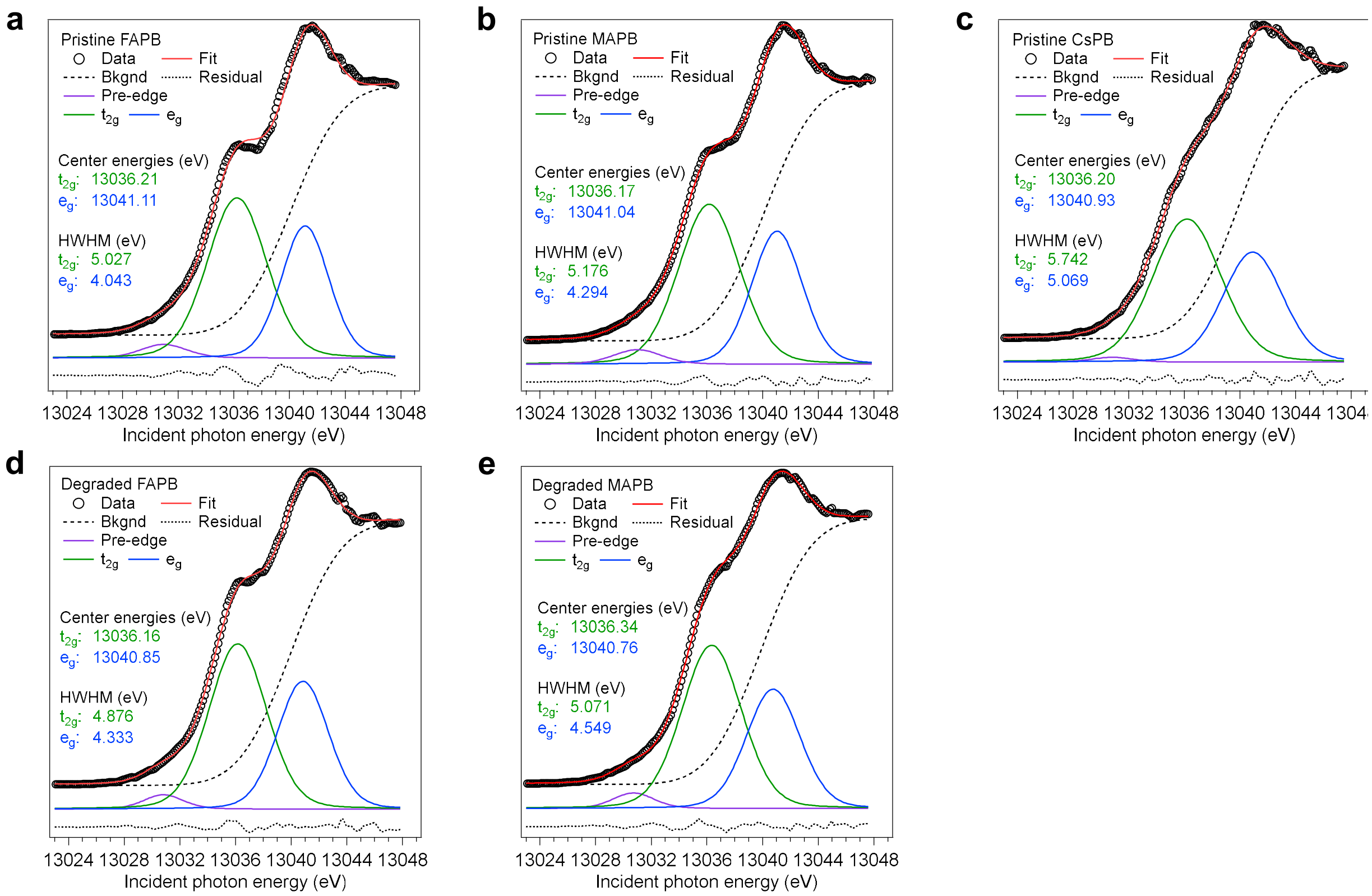

Supplementary Fig. 7. **Curve-fitted lead $L_3$ main-edge spectra.**

Supplementary Table 1. **Summary of curve-fitted parameters extracted from bromine $K\beta_1$ X-ray emission spectra of pristine and photodegraded APB and $PbBr_2$.**

| Compound | Pristine | Degraded |
|---|---|---|
| FAPB | Peak position: 13284.368 ± 0.003 eV<br><br>Voigt FWHM: 4.577 eV<br><br>Gauss width: 1.780 ± 0.003 eV<br><br>Lorenz width: 3.884 ± 0.008 eV<br><br>Shape factor: 1.8162 | Peak position: 13284.384 ± 0.003 eV<br><br>Voigt FWHM: 4.607 eV<br><br>Gauss width: 1.867 ± 0.004 eV<br><br>Lorenz width: 3.850 ± 0.007 eV<br><br>Shape factor: 1.7173 |
| MAPB | Peak position: 13284.627 ± 0.003 eV<br><br>Voigt FWHM: 4.60 eV<br><br>Gauss width: 1.844 ± 0.004 eV<br><br>Lorenz width: 3.866 ± 0.008 eV<br><br>Shape factor: 1.7458 | Peak position: 13284.540 ± 0.003 eV<br><br>Voigt FWHM: 4.55 eV<br><br>Gauss width: 1.842 ± 0.004 eV<br><br>Lorenz width: 3.809 ± 0.008 eV<br><br>Shape factor: 1.7214 |
| CsPB | Peak position: 13284.695 ± 0.003 eV<br><br>Voigt FWHM: 4.57 eV<br><br>Gauss width: 1.811 ± 0.004 eV<br><br>Lorenz width: 3.852 ± 0.009 eV<br><br>Shape factor: 1.771 | Peak position: 13284.539 ± 0.003 eV<br><br>Voigt FWHM: 4.55 eV<br><br>Gauss width: 1.584 ± 0.003 eV<br><br>Lorenz width: 3.999 ± 0.008 eV<br><br>Shape factor: 2.1014 |
| $PbBr_2$ | Peak position: 13285.315 ± 0.002 eV<br><br>Voigt FWHM: 4.50 eV<br><br>Gauss width: 1.475 ± 0.002 eV<br><br>Lorenz width: 4.014 ± 0.005 eV<br><br>Shape factor: 2.2658 | |

Supplementary Note 1. **Types of information yielded by X-ray absorption and emission spectroscopy.**

As multi-elemental compounds, HaPs are well-suited for study with core level spectroscopies which offer element-, orbital- and crystal symmetry-selectivity [55].

We use an energy-dispersive hard X-ray spectrometer to perform a range of concurrent measurements during photodegradation, for each element [53]. For bromine, we probe the relative bromine-lead bond ionicity via bromine $K\beta_1$ X-ray emission spectroscopy (XES), bromine *p*-projected density of states (*p*-PDOS) in the conduction band via bromine *K*-edge (*1s*→*np* where $n \geq 4$) High Energy Resolution Fluorescence Detected X-ray Absorption Spectroscopy (HERFD-XAS) and Br *4p*-PDOS in the valence band via bromine valence-to-core (VtC) XES [29,53]. For lead, we use lead $L_3$-edge ($2p_{3/2}$→*ns*,*nd* transition where $n \geq 6$) HERFD-XAS to probe the internal structure and degree of cooperative tilting of the $PbBr_6$ octahedra and the lead *s*-/*p*-/*d*-PDOS in the conduction band. Few reports currently exist regarding the application of HERFD-XAS to HaPs [29,35,56]. The HERFD variant of XAS, which requires an X-ray emission spectrometer, is particularly crucial for the lead $L_3$-edge as the intrinsic core-hole lifetime broadening can be reduced from 6.1 eV to 2.5 eV [57,58]. Previously, we were able to discern A-cation-induced changes in the crystal structure which are reflected in the lead $L_3$ HERFD spectra; the conventional total fluorescence yield (TFY) spectra resemble mostly featureless sigmoidal step functions [29]. We optimized the measurement conditions here to yield comparable spectrometer runtimes for bromine *K*-edge and lead $L_3$-edge HERFD-XAS, enabling us to track potential chemical bonding, crystal structure and electronic structure changes as a function of beam exposure in parallel.

Reported ground-state electronic structure calculations indicate that the states near the CBM are comprised of hybridized bromine *4p*-lead *6p* states [59,60]. We directly probe the bromine *4p* component of the hybridized states near the CBM via bromine *1s*→*4p* XAS. Lead $2p_{3/2}$→*6s*/*6d* XAS normally does not probe lead *6p* states, due to dipole selection rules, but Drisdell *et al*. have found that electronic transitions to hybridized Pb *d*/*p*-Br *p* states contribute to the pre-edge region of the lead $L_3$ spectrum of MAPB [35]. Hence, lead $L_3$-edge XAS offers a complementary lead-based view of the hybridized Br *4p*-Pb *6p* states near the CBM. To connect potential changes in the spectra with changes in the material, the assignment of electronic transitions in the spectra using XAS calculations is required. This has been performed for the lead $L_3$-edge by Drisdell *et al.* [35] and for the bromine *K*-edge by us [29]. Reported electronic structure calculations indicate that the states near the VBM are comprised of hybridized bromine *4p*-lead *6s* states, where the bromine *4p* states contribute the highest intensity [59]. We probe the distribution of bromine *4p* states in the valence band with bromine VtC XES.

Supplementary Note 2. **Checks for decomposition into binary precursor compounds.**

The photodegraded bromine *K* and lead $L_3$ spectra of the all-inorganic APB, CsPB (Fig. 1g,h), resemble its pristine spectra with an energetic shift, indicating its unoccupied bromine- and lead-projected electronic structure is negligibly affected by photodegradation but some material change has occurred. The photodegraded spectra of the hybrid organic APB's (Fig. 1a,b,d,e) bear some resemblance to its pristine spectra. We have previously reported that bromine *K*-edge XAS of APB's probes two energetically separated bromine *4p* distributions of states in the main-edge, the σ (~13466 to ~13470 eV) and π (~13470 to ~13475 eV) states, which reflects the two bonding characters of the bromine *4p*-PDOS in the conduction band. The energetic separation between the peak intensities of these two distributions is the σ-π energy separation and its magnitude is influenced by the strength of N-H…Br hydrogen bonding [29]. Photodegradation of the organic APB's has shifted the absorption onsets and appears to have increased the σ-π energy separation (Fig. 1a,d).

We have plotted the bromine *K* spectra of the binary precursor compounds ABr and $PbBr_2$, and 1 : 1 averaged ABr : $PbBr_2$ for comparison to the spectra of the photodegraded organic APB's. We have performed beam damage checks on ABr and $PbBr_2$ (Supplementary Fig. 2), the spectra of pristine ABr and $PbBr_2$ are displayed in Supplementary Fig. 3, and the averaging of the ABr : $PbBr_2$ spectra is shown in Supplementary Fig. 4. The bromine *K* spectra of photodegraded MAPB/FAPB do not overlap well with the ABr : $PbBr_2$ spectrum. Rather, the energetic alignment via shifting of the photodegraded spectra with respect to the pristine spectra, at the onset and at the highest-energy portion of the main-edge, offers a better match (Supplementary Fig. 6a,b,d,e), supporting the existence of the widening of the σ-π energy separation. As a second check for decomposition, we compare the photodegraded lead $L_3$ spectra of the organic APB's to $PbBr_2$ and metallic lead (foil) (Fig. 1b,e). The spectra do not overlap well; the photodegraded spectra appear as shifted and broadened versions of the pristine spectra. As a third check for decomposition into $PbBr_2$, we have examined the change in the bromine $K\beta_1$ peak position upon photodegradation (Supplementary Fig. 1). The peak position shifts from 13284.63 to 13284.54 for MAPB and is roughly unchanged for FAPB (13284.37 eV). The peak position for $PbBr_2$ is 13285.32 eV. The spectrometer energy uncertainty is ± 0.020 eV and the $K\beta_1$ Voigt fit uncertainty is ± 0.004 eV. Photodegradation of MAPB has caused its Pb-Br bond to become more ionic, instead of more covalent like the Pb-Br bond in $PbBr_2$. As a fourth check, we have observed the absence of asymmetry in the $K\beta_1$ XES peaks which indicates that a single bromine-containing composition, not multiple compositions, is present in the sample region probed. Consequently, we conclude that $APbBr_3 \rightarrow ABr + PbBr_2$ decomposition, if present, is negligible in the sample regions probed and the photodegraded APB's remain perovskites.

Supplementary Note 3. **Lead $L_3$ main-edge curve fitting procedure.**

Since the Pb $L_3$ HERFD-XAS spectrum of MAPB is comprised of transitions to a range of hybridized states, including Pb *d*/*p*-Br *p* hybridized states, our curve-fitting, which relies on an oversimplified assignment of the rising- and main-edge features to transitions to Pb *d*-Br *d* hybridized states (which account for the strongest excitations), provides semi-quantitative information [35]. X-ray absorption spectroscopy curve fitting was performed with version 0.9.26 of the Athena XAS fitting program [61]. To extract meaningful trends from potential changes in the center energy and/or energetic width (full width half maximum (FWHM)) of the $t_{2g}$ and $e_g$ spectral features, we used four types of curve fitting constraints.

First, the minimum number of line shapes, four, is used for the fits of all spectra: an error function for the background and three pseudo-Voigt peaks for the onset/pre-edge, $t_{2g}$ and $e_g$ features. Pseudo-Voigt peak fits are used since the Lorentzian broadening originates from the core hole lifetime and the Gaussian broadening originates from the instrumentation (e.g. beamline monochromator).

Second, the type (error function) and width (4.10 eV) of the background line shape are fixed for all curve fitted spectra. The initial center energy (e0) of the error function background is derived from the curve fit of the spectrum of pristine CsPB (13039.96 eV). Background center energies for the other spectra are shifted based on shifts of the onsets of the other spectra relative to the onset of the spectrum of pristine CsPB.

Third, the pseudo-Voigt Gaussian-Lorentzian-mixing parameter (eta) for the fitted $t_{2g}$ and $e_g$ features is fixed at 0.2. The pseudo-Voigt mixing parameter for the fitted onset/pre-edge feature is fixed at 0.3.

Fourth, the energy separation between the center energies of the error function background and the onset pseudo-Voigt peak is fixed at 9.14 eV.

The eight unconstrained/fitted parameters are the center energies, FWHM and areas of the $t_{2g}$ and $e_g$ pseudo-Voigt peaks and FWHM and area of the onset/pre-edge pseudo-Voigt feature. The upper bound of the energy uncertainty of the fitted FWHM is 40 meV.

We assess the reliability of the fits both quantitatively and subjectively. The $t_{2g}$ : $e_g$ area ratio is expected to be 3 : 2 (from the five degenerate d-states), and is indeed ~3 : 2 for all of the fits. The energy uncertainty of the fitted center energy of the $t_{2g}$ feature ranges from 0.1 – 0.4 eV and is of comparable magnitude to the ~0.3 eV photon bandwidth of the Si(311) beamline monochromator. The center energy of the $t_{2g}$ feature for the pristine APB compounds is the same, at 13036.2 ± 0.3 eV, which is expected since the three pristine APB compounds contain the same type of $PbBr_6$ octahedra. The maximum peak-to-peak amplitudes of the residuals is comparable to the peak-to-peak noise of the measured $e_g$ features.

Supplementary Note 4. **Lead $L_3$ difference spectra of continuously photodegrading MAPB.**

We generate lead $L_3$ difference spectra by subtracting the spectra for spectrometer runs 3 to 13 from the pristine spectrum (Fig. 2c). We observe that the difference spectra show the evolution of spectral features in three energy regions: < 13036 eV, ~13036 to ~13040 eV, and > 13040 eV, which we refer to as regions D, E and F, respectively. Region D is predominantly sensitive to shifts of the onset, region E is predominantly sensitive to changes of the width of $e_g$ and region F is predominantly sensitive to changes of the center energy and/or width of $e_g$. Based on the temporal evolution of the relative intensities of the features in regions D, E and F, we group the spectrometer runs into three temporal regimes that we label as regimes I, II and III, respectively. Regime I, comprising run 3, shows negligible changes in all energy regions. Regime II, encompassing runs 4 to 10, show negative intensity in region D (onset shifted to higher energy), gradually increasing positive intensity in region E (gradual broadening of $e_g$) and negative intensity in region F (decrease of crystal field splitting). Regime III, encompassing runs 11 to 13, shows noticeably higher negative intensity in region D (larger onset shift relative to regime II) and comparable intensities in regions E and F to the last run from regime II. We generate representative spectra from regimes II and III and plot them in Fig. 2d, along with the labels for the three energy regions D, E and F.